\documentclass[reprint,aps, pre]{revtex4-1}
\usepackage[margin=1in]{geometry}

\usepackage{hyperref} 
\usepackage{xurl}

\usepackage{varwidth}

\usepackage{subcaption}
\usepackage{mathtools, amsmath, amssymb, graphicx, amsfonts}
\usepackage{xcolor, comment} 
\usepackage{bbm} 
\usepackage{float} 
\usepackage{amsopn} 
\usepackage{mathrsfs} 
\usepackage{algorithm} 
\usepackage{algorithmic} 
\usepackage{float}
\usepackage{flushend}
\usepackage[aboveskip=4pt]{subcaption}
\usepackage{float}

\usepackage{tcolorbox}
\usepackage{framed}

\begin{document}

\title{Bounded-Confidence Models of Opinion Dynamics with Neighborhood Effects}
\author{Sanjukta Krishnagopal}
\email{sanjukta@ucsb.edu}
\affiliation{Department of Computer Science, University of California, Santa Barbara}
\author{Mason A. Porter}
\email{mason@math.ucla.edu}
\affiliation{Department of Mathematics, University of California, Los Angeles;}
\affiliation{Department of Sociology, University of California, Los Angeles;}
\affiliation{Santa Fe Institute}


\begin{abstract}
People are influenced by the people with whom they interact, and their social networks evolve as their opinions change. In this paper, we generalize bounded-confidence models (BCMs) of opinion dynamics by incorporating neighborhood effects.
In a BCM, interacting agents influence each other through dyadic influence if their opinions are sufficiently similar to each other. In our ``neighborhood BCMs'' (NBCMs), interacting agents {are influenced both by} each other's opinions and {by} the opinions of the agents in each other's neighborhoods. Our NBCMs thus include both the usual dyadic influence between agents and a ``transitive influence'', which encodes the influence of an agent's neighbors, when determining whether or not an interaction changes the opinions of agents. In this transitive influence, an individual's opinion is influenced by a neighbor when, on average, the opinions of the neighbor's neighbors are sufficiently similar to its own opinion. We formulate both neighborhood Deffuant--Weisbuch (NDW) and neighborhood Hegselmann--Krause (NHK) BCMs. We simulate our NDW model on a variety of types of networks, and we observe interesting opinion dynamics, such as opinion jumping and pseudo-consensus, that cannot occur in the standard DW model.

We build further on our NBCMs by introducing a neighborhood-based network adaptation in which a network coevolves with agent opinions by changing its structure through ``transitive homophily". In this network evolution, an agent breaks a tie to one of its neighbors and then rewires that tie to a new agent, with a preference for agents with a mean neighbor opinion that is closer to its own opinion. Using numerical simulations on a variety of types of networks, we explore how the qualitative opinion dynamics and network properties of our adaptive NDW model change as we adjust the relative proportions of dyadic and transitive influence. In our numerical experiments, we find that incorporating neighborhood effects into the opinion dynamics and the network-adaptation rewiring strategy tends to reduce the spectral gap and degree assortativity of networks. 

\end{abstract}


\maketitle



\section{Introduction} \label{intro}

Human opinions, human behavior, and social interactions influence each other in inextricable ways. Human opinions and their dynamics play an important role in many real-world settings, such as decision-making~\cite{milburn1991persuasion}, misinformation and disinformation campaigns \cite{guo2022effect}, referendums \cite{lin2022online}, and political polarization \cite{jost2022cognitive}. {A major factor in all of these realms is homophily, which encapsulates the tendency of individuals to interact more with individuals that are similar to themselves than with those {that} differ from them \cite{mcpherson2001birds}.}

There are myriad models of opinion dynamics \cite{xia2011opinion,noorazar2020}, which encode opinions and how they evolve. Some opinion models (e.g., voter models \cite{redner2019}) have discrete-valued opinions, and other opinion models have continuous-valued opinions. Models with continuous-valued opinions include the DeGroot consensus model \cite{french1956formal,degroot1974reaching}, the Friedkin--Johnson model \cite{friedkin1990social}, bounded-confidence models (BCMs) \cite{lorenz2007continuous,bernardo2024}, and models with explicit radicalization dynamics~\cite{baumann2020modeling}. Much of the literature on opinion dynamics has focused on conditions for consensus, in which {the opinions of a network of agents} converge to a single value, but it is also important to study situations such as polarization (in which there are two major opinion clusters), fragmentation (in which there are three or more major opinion clusters), and others.

In a traditional BCM, interacting agents change their opinions if and only if their current opinions are sufficiently similar. Specifically, two agents influence each other if the difference between their opinions lies within a confidence bound. The two foundational BCMs are the Hegselmann--Krause (HK) model (which has synchronous updates of agent opinions) \cite{rainer2002opinion} and the Deffuant--Weisbuch (DW) model (which has asynchronous opinion updates) \cite{weisbuch2002meet}. In the HK model, interactions take place in groups, with each agent changing its opinion to the mean of the opinions of its neighboring agents whose opinions are within the confidence bound. By contrast, in the DW model, interactions are pairwise (i.e., dyadic), with a single pair of agents interacting at each discrete time. If the difference between {the opinions of two interacting individuals} lies within the confidence bound, they compromise their opinion by some amount. Extensions of BCMs include models that incorporate leaders \cite{zhao2016bounded}, stubborn agents \cite{tian2018opinion}, heterogeneous confidence bounds \cite{sobkowicz2015extremism}, smooth interactions (e.g., through sigmoidal functions) \cite{brooks2023}, coevolving opinions and networks \cite{brede2019does,kan2023}, and polyadic agent interactions (i.e., with three or more agents interacting simultaneously)~\cite{hickok2022bounded, schawe2022higher}.

By studying BCMs on networks, researchers examine how network structure influences opinion dynamics \cite{meng2018}. Most BCMs incorporate only direct influence, with interacting agents considering only each other's opinions when determining whether or not to update their opinion. However, humans are influenced not only directly but also through transitive influence (e.g., through friends of friends) \cite{liu2011trust}. They are thus influenced by the neighbors of adjacent agents. Such transitive influence is notably prevalent in interactions on social media~\cite{christakis2009connected}, which has low barriers to interacting with friends of friends. Moreover, individuals can be influenced not only by the friends of their friends (i.e., individuals {that} are two steps away), but also by individuals that are even farther away in a network. In their studies of influence on social networks, Christakis and Fowler \cite{christakis2013} posited the idea of ``three degrees of influence" to illustrate how many steps away from an agent in a network can still yield robust influence on human behavior and opinions. Along similar lines, Miranda et al.~\cite{miranda2024indirectsocialinfluencediffusion} studied innovation adoption and observed that the adoption rate of an innovation is influenced significantly both by nearest neighbors and by extended neighborhoods. Researchers have also highlighted the importance of extended neighborhoods in empirical network data (e.g., in Facebook \cite{backstrom2011}) and in the design of algorithms (e.g., for local community detection \cite{jeub2015}).

In the present paper, we study opinion models that incorporate transitive influence from neighbors of neighbors. We formulate \emph{neighborhood BCMs} (NBCMs), in which an agent can change its opinion based both on dyadic interactions with adjacent nodes and on the mean opinion of the neighbors of those adjacent nodes. We refer to this mean opinion as the ``mean neighbor opinion". For the dyadic influence, we employ the usual confidence bound of BCMs, so two interacting agents compromise their opinions by some amount if they are sufficiently similar to each other (i.e., if their opinions differ by less than a confidence bound). The two agents also experience transitive influence. In such influence, an agent is influenced by a neighboring agent with whom it interacts if {the neighboring agent's} mean neighbor opinion is within the confidence bound (i.e., if the mean opinion of its neighboring nodes is sufficiently similar). Although the mean neighbor opinion influences whether or not an agent updates its opinion, the magnitude of opinion changes are affected only by the opinion of the interacting agents themselves. We explore the qualitative behavior of our NBCMs for different relative proportions of direct influence (i.e., through dyadic interactions) and transitive influence (i.e., through neighborhood effects). Because of the transitive influence (and unlike in conventional BCMs), a node in our NBCMs can update its opinion due to an interaction with an adjacent node without requiring that the adjacent node also update its opinion, even when both nodes have the same confidence bound. In particular, this asymmetry can arise when all nodes have the same constant confidence bound. Additionally, because of transitive influence, two interacting nodes can influence each other even when their opinions differ by more than the confidence bound.

We are not aware of any existing BCMs that consider the above neighborhood effects. However, there does exist a DeGroot model in which agents update their opinions so that they become closer to the opinions of agents that are neighbors of their neighbors~\cite{zhou2020two}, and a very recent paper incorporated neighborhood effects into a voter model \cite{munoz2024}. The neighborhood effects in our NBCMs model ``transitive homophily" (i.e., neighborhood-based homophily) and differ in a key way from the neighborhood effects in \cite{zhou2020two}. In our model, changes in agent opinions are based only on the opinions of their neighbors (i.e., by agents that are one step away in a network and hence are adjacent to them). 
However, neighbors of neighbors (i.e., agents that are two steps away) influence whether or not an agent changes its opinion in the first place. Our NBCMs also inherit the desirable property that individuals have heterogeneous tolerances towards others' opinions. Notably, this heterogeneity is present even for a constant, homogeneous confidence bound. 

{We also study an NBCM that coevolves with network structure.
We consider \emph{neighborhood-based adaptation}, in which an agent can form or sever connections with another agent based both on the similarity of the other agent's opinions to their own and on the mean opinion of the neighbors of that other agent.}
Changes in opinions can lead to changes in relationships between agents, which in turn can lead to changes in the structure of a network. See \cite{berner2023} for a review of coevolving (i.e., adaptive) network models. A variety of adaptive-network models have been developed to study the coevolution of opinions and network structure. Examples include adaptive voter models~\cite{holme2006nonequilibrium,durrett2012}, adaptive BCMs and related opinion models~\cite{brede2019does,del2017modeling,kozma2008consensus,parravano2016bounded,kan2023}, and others. In an adaptive opinion model, there is typically a mechanism for agents to remove ties to agents with whom they disagree (or with whom they disagree sufficiently) and then establish ties to other agents, perhaps (for convenience) in a way that preserves the density of ties (i.e., edges). Our adaptive NBCM has two notions of homophily: (1) \emph{dyadic homophily}, in which agents with more similar opinions are more likely to have social ties with each other (i.e., ``birds of a feather flock together"); and (2) \emph{transitive homophily}, in which agents are more likely to be influenced by agents whose mean neighbor opinion is closer to their own opinion (i.e., ``you are who you know"). We employ an edge-rewiring strategy that is based on transitive homophily and is similar to the strategy in \cite{kan2023}. When an agent breaks an edge, it rewires this edge to another agent with a probability that is proportional to the similarity between its opinion and the mean neighbor opinion. This rewiring strategy explicitly considers the neighborhoods of nodes. Given network adaptation through transitive homophily, we study how network structure and opinions coevolve for different relative importances of dyadic influence (which arises from opinion similarity between adjacent agents) and transitive influence (which arises from opinion similarity of an agent with the mean neighbor opinion of an adjacent agent). We quantify this relative importance through a neighborhood-tuning parameter. Our NBCM simulations reveal unexpected and nonmonotic dependence of various network measures (such as degree assortativity and the adjacency matrix's spectral gap) on the neighborhood-tuning parameter.

Our paper proceeds as follows. In Section \ref{sec:BCMs}, we discuss BCMs on networks. In Section \ref{sec3}, we present our NBCMs both without and with network adaptation. In Section \ref{sec:nbc}, we introduce our NBCMs. We formulate both a neighborhood DW (NDW) model and a neighborhood HK (NHK) model. In Section \ref{sec: rewiring_opinions}, we incorporate an edge-rewiring mechanism for network adaptation into our NDW and NHK models. In this rewiring mechanism, the mean neighbor opinions of nodes affect the establishment and severing of edges. In Section \ref{sec:opinion_empirics}, we examine the behavior of our adaptive NDW model. We present the results of simulations of our {non-adaptive} NDW model on time-independent networks in Section \ref{oppy}, present the results of simulations of our adaptive NDW model in Section \ref{sec:op_network}, discuss the effects of parameters on opinion dynamics in Section \ref{new-c}, discuss simulations of our adaptive NDW model with homophilic rewiring on different types of networks in Section \ref{new-d}, and examine the effects of network size on the qualitative dynamics of our adaptive NDW model in Section \ref{new-e}. In Section \ref{conclusions}, we summarize our results and discuss various extensions of our {work}. Our code is available at \url{https://bitbucket.org/neighborhood-bounded-confidence-model-of-opinion-dynamics}.


\section{Bounded-confidence models (BCMs) on networks}\label{sec:BCMs}

In a BCM, an agent is receptive to the opinions of another agent if and only if the opinions of these two agents are sufficiently similar (i.e., they lie within some confidence bound) \cite{lorenz2007continuous,bernardo2024}.
Consider an unweighted and undirected network (i.e., graph) $G$ with $N$ agents. Each agent $i$ is a node of the network. We denote the network's associated adjacency matrix by $A$, where $A_{i,j} = 1$ if there is an edge between agent $i$ and agent $j$ and $A_{i,j} = 0$ if there is no edge between them. We {assume} that the network is unweighted for simplicity, but we can consider weighted networks by letting $A_{i,j}$ be the weight of the edge between nodes $i$ and $j$. We refer to a pair of nodes as a ``dyad". Suppose that each agent $i$ has a continuous-valued opinion $x_i(t) \in [0,1]$ at discrete time $t$. The vector of opinions of the $N$ agents is the ``opinion profile'' $X(t) = (x_1(t),\ldots, x_N(t))$ at time $t$. A subset of nodes such that the opinions of consecutive nodes (which we order by their opinion values) {are all} within $\epsilon$ of each other is an ``opinion cluster''. An opinion cluster is in ``consensus" when all of its nodes have the same opinion. However, it is also possible for an opinion cluster to be in a different state, such as a pseudo-consensus (see Section \ref{oppy}).
 
In the DW model, opinion updates are asynchronous. At each discrete time, one chooses an edge uniformly at random; the nodes that are attached (i.e., incident) to that edge potentially compromise their opinions. Suppose that nodes $i$ and $j$ interact at time $t$. They update their opinions according to the rule
\begin{align} \label{eq:DW}
	x_i (t + 1) &= x_i(t) + \rho (x_j(t) - x_i(t)) \mathbbm{1}_{d_{\text{BC}}(i,j) < \epsilon_{i,j}} \,, \notag \\
	x_j (t + 1) &= x_j(t) + \rho (x_i(t) - x_j(t)) \mathbbm{1}_{d_{\text{BC}}(i,j) < \epsilon_{i,j}}\,, 
\end{align}
where the distance between the opinions of agents $i$ and $j$ is $d_{\text{BC}}(i,j) = |x_i(t) - x_j(t) |$, the indicator function $\mathbbm{1}_b = 1$ if condition $b$ holds and $\mathbbm{1}_b = 0$ if condition $b$ does not hold, $\epsilon_{i,j}$ is the confidence bound, and $\rho \in (0,0.5]$ is a constant (which is sometimes called a ``convergence parameter"). The value $\rho = 0.5$ corresponds to an exact opinion compromise; in this case, two interacting agents that update their opinions both adopt the mean of their opinions. The confidence bound $\epsilon_{i,j}$ is symmetric in most studies, but one can make it asymmetric (i.e., $\epsilon_{i,j} \neq \epsilon_{j,i}$) if one desires. For a symmetric confidence bound, the opinion updates of nodes $i$ and $j$ are symmetric because $d_{\text{BC}}(i,j) = d_{\text{BC}}(j,i)$. For simplicity, we suppose that the confidence bound is homogeneous, so $\epsilon_{i,j} = \epsilon$ for all $i,j \in \{1,2, \ldots, N\}$. Therefore, when we discuss a confidence bound in the rest of our paper, we refer to it as ``the" confidence bound. When adjacent nodes have opinions that differ by less than the confidence bound, we say that these nodes are ``directly receptive" to each other.

In the HK model, opinion updates are synchronous and hence deterministic. At each discrete time, an agent interacts with all of its neighboring agents. Each agent updates its opinion to the mean of the opinions of all neighbors with opinions within the confidence bound. That is,
\begin{equation}
	x_i (t + 1) = x_i(t) + \frac{\rho}{| \Gamma_i|}\sum_j A_{i,j} x_j(t) \mathbbm{1}_{d_{\text{BC}}(i,j) < \epsilon)} \,,
\label{eq:HK}
\end{equation}
where ${\Gamma}_i = \{j | A_{i,j} = 1 \quad \text{and} \quad |x_i - x_j| < \epsilon \}$ is the set of adjacent agents (i.e., neighbors) of agent $i$ whose opinions are within the confidence bound. 
Given a network and an initial opinion profile, the HK model is deterministic. This is convenient for mathematical analysis.

Consider the HK model on a connected network (i.e., a network in which there exists a path from any node to any other node) with a homogeneous confidence bound $\epsilon$. {In matrix form, Eq.~\eqref{eq:HK} is ${X(t + 1)} = W(t)X(t)$, where the opinion-update matrix $W$ is related to the adjacency matrix and $X(t)$ is an opinion profile. If the matrix $W$ is row-stochastic and type-symmetric (i.e., $\mathrm{sign}(W_{i,j}) = \mathrm{sign}(W_{j,i})$ and $W_{i,j} = 0$ if and only if $W_{j,i} = 0$) with nonzero diagonal entries and non-negative entries (which is true of the adjacency matrix $A$), then the HK model has the following properties~\cite{rainer2002opinion,lorenz2005stabilization}}:
\begin{itemize}
	\item{Opinion updates do not change {the order of the node} opinions. That is, $x_i(t) \leq x_j(t)$ at any time $t$ implies that $x_i(\tau) \leq x_j(\tau$) for all times $\tau > 0$. See Section 3D of \cite{rainer2002opinion}.}
	\item{If the opinion difference between two nodes exceeds the confidence bound $\epsilon$ at time $t$, those two nodes cannot subsequently have opinions that differ by less than the confidence bound. See Section 3D of \cite{rainer2002opinion}.}
	\item{A necessary condition to achieve a consensus opinion, which entails that $|x_i(T) - x_j(T)| \rightarrow 0$ for all nodes $i$ and $j$, in a finite time $T$ is that removing all edges between adjacent nodes that are not directly receptive to each other (i.e., neighboring nodes whose opinions differ by at least the confidence bound) yields a connected pruned network. If the pruned network is disconnected at time $t$, it remains disconnected at all subsequent times.}
	\item{By Theorem 2 of \cite{lorenz2005stabilization}, the opinion profile converges to an opinion profile with at least one opinion cluster and consensus in each opinion cluster.}
\end{itemize}

These properties are nice mathematically, but they differ markedly from typical real-life observations of opinion dynamics. Additionally, the opinion updates in the HK and DW models do not incorporate neighborhood effects. In Section \ref{sec:nbc}, we generalize the HK and DW models to incorporate such effects. The resulting neighborhood-based HK model has rather different properties than the ones that we just discussed. Introducing neighborhood effects also yields interesting phenomena in DW models.


\section{Neighborhood-based opinion models and network adaptation}\label{sec3}


\subsection{Neighborhood bounded-confidence models}
\label{sec:nbc}

We introduce neighborhood bounded-confidence models (NBCMs) that generalize the HK and DW models. Incorporating neighborhood effects allows us to investigate how an agent's proclivity to change its opinion depends both on the opinions of its adjacent agents and on the opinions of the neighbors of its adjacent agents. This idea encompasses the notion that ``you are who you know". For example, consider a mildly liberal agent $i$ that is adjacent to two agents, $j$ and $k$, that have the same centrist political opinion as each other but have different neighborhoods. Suppose that most nodes in $j$'s neighborhood are liberals but that most nodes in $k$'s neighborhood are conservatives. It seems plausible that agent $i$ is influenced differently by nodes $j$ and $k$, perhaps with greater influence from {node} $j$ because its neighborhood has a political opinion that is closer to $i$'s opinion. We extend both the HK and DW models to incorporate such neighborhood effects.

\begin{figure*}[htbp!]
		\includegraphics[width=\linewidth]{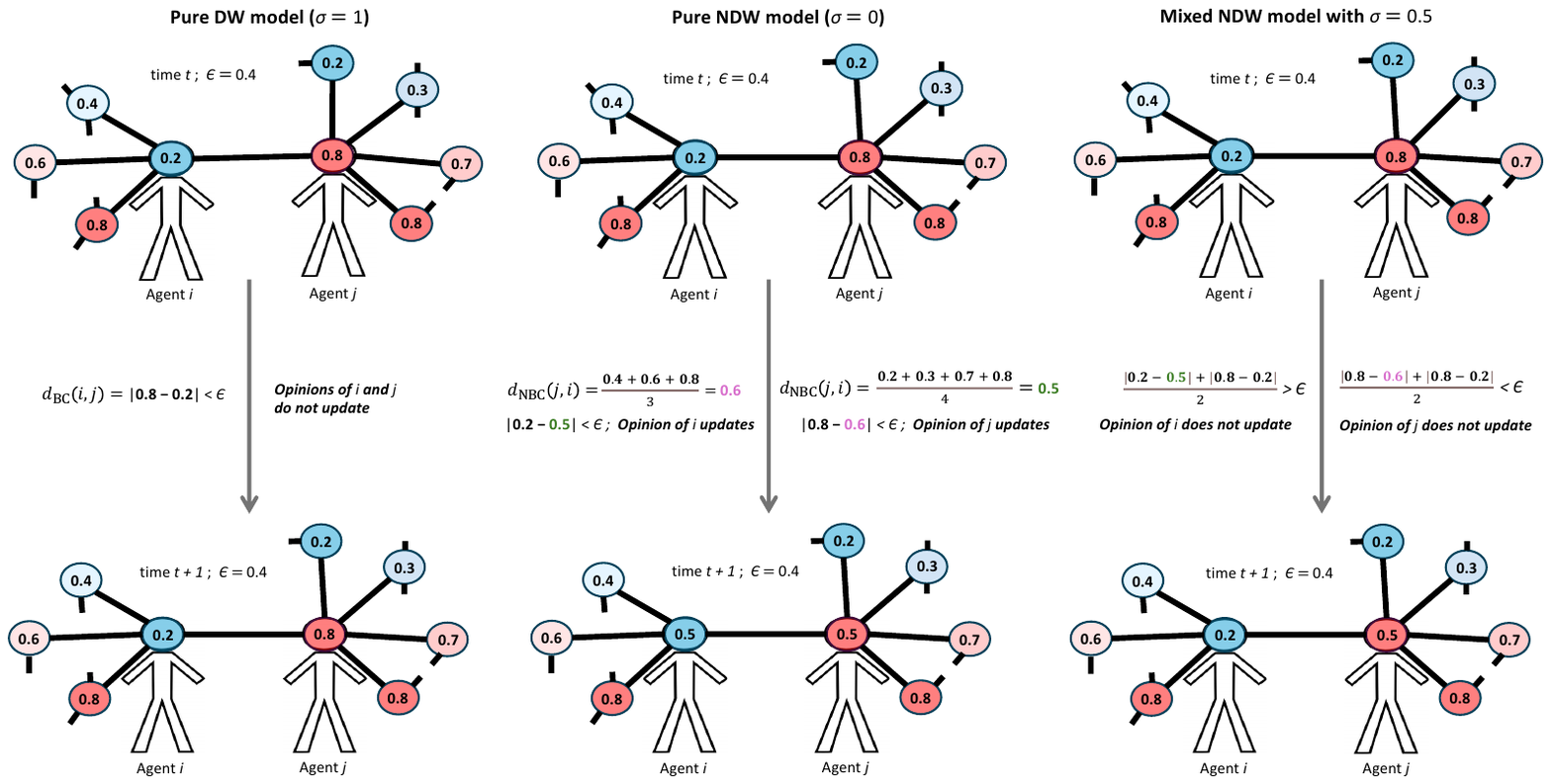}
	\vspace{-2.5 cm}
	\caption{A schematic illustration of opinion updates of two adjacent nodes, $i$ and $j$, in the pure (i.e., baseline) DW model, pure NDW model, and the NDW model with neighborhood-tuning parameter $\sigma = 0.5$ (i.e., a ``mixed" NDW model). All edges have unit weight and convergence parameter $\rho = 1$.}
	\label{fig:example}
\end{figure*}

In our neighborhood DW (NDW) model, opinion updates are asynchronous. At each discrete time $t$, we choose $f = nN$ edges uniformly at random without replacement. If $nN$ is not an integer, we take $f = \lceil nN \rceil$ {(}where $\lceil \cdot \rceil$ is the ceiling function) to round up to the nearest integer. (For all reported numerical computations, $nN$ is an integer.) The two incident nodes of each edge interact dyadically. We perform opinion updates sequentially; a node that updates its opinion uses its new opinion for subsequent interactions at time $t$.

Suppose that $i$ and $j$ are the two nodes that are attached (i.e., incident) to a chosen edge. They update their opinions according to the rule
 \begin{widetext}
\begin{align} 	\label{eq:NBC_DW}
	x_i (t + 1) &= x_i(t) + \rho (x_j(t) - x_i(t))  \mathbbm{1}_{\sigma d_{\text{BC}}(i,j) + (1 - \sigma) d_{\text{NBC}}(i,j) < \epsilon} \,, \notag \\
	x_j (t + 1) &= x_j(t) + \rho (x_i(t) - x_j(t))  \mathbbm{1}_{\sigma d_{\text{BC}}(i,j) + (1 - \sigma) d_{\text{NBC}}(j,i) < \epsilon} \,, 
\end{align}
\end{widetext}
where the distance $d_{\text{NBC}}(i,j) = |x_i(t) - \frac{\sum_k A_{jk} w_k x_k(t)}{|\Gamma_j|}|$ and $w_k$ is the weight of node $k$. Node weights can take a variety of possible values, and heterogenous node weights can capture ideas like heterogeneous importances or activity levels~\cite{li2022bounded}{. We} suppose for simplicity that $w_k = 1$ for all $k$. The \textit{neighborhood-tuning parameter} $\sigma \in [0,1]$ controls the relative weights for an agent $i$ to compromise its opinion with the opinion of its neighbor $j$ due to (1) the similarity between {node $i$'s opinion and node $j$'s opinion} and (2) the similarity between node $i$'s opinion and node $j$'s mean neighbor opinion~\footnote{Node $i$'s opinion contributes to this mean.} (i.e., a neighborhood-based transitive influence). With $\sigma = 1$, we recover the baseline DW model \eqref{eq:DW}. With $\sigma = 0$, we obtain a ``pure'' NDW model \eqref{eq:NBC_DW}. Intermediate values of $\sigma$ combine the influence of opinion similarity between individual agents and {neighborhood-based opinion similarity}. We sometimes refer to this situation as a ``mixed" NDW model. The convergence parameter $\rho$ controls the amount that interacting agents compromise in one opinion-update step. The value $\rho = 0.5$ corresponds to an exact opinion compromise. The distance $d_{\text{BC}}$ depends symmetrically on agents $i$ and $j$, but the distance $d_{\text{NBC}}$ depends asymmetrically on the two agents. Agent $i$ looks at agent $j$'s neighbors to determine whether or not it is influenced \footnote{In some prior studies, such as in investigations of DeGroot models \cite{xia2015structural,liu2022opinion}, researchers have interpreted influence as a notion of trust. With such an interpretation, the amount of trust that agent $i$ places in agent $j$ depends on $j$'s neighborhood.} by $j$, whereas agent $j$ looks at agent $i$'s neighbors to determine whether or not it is influenced by $i$. In Fig.~\ref{fig:example}, we show a schematic illustration of opinion updates (with exact opinion compromises) between two interacting nodes for the pure (i.e., baseline) DW model, the pure NDW model, and a mixed NDW model with $\sigma = 0.5$. 

In our neighborhood HK (NHK) model, opinion updates are synchronous. The opinions of the nodes update according to the equation
 \begin{widetext}
\begin{equation}
	x_i(t + 1) = x_i(t) + \frac{\rho}{| \Gamma_i|} \sum_j A_{i,j} x_j(t)  \mathbbm{1}_{\sigma d_{\text{BC}}(i,j) + (1 - \sigma) d_{\text{NBC}}(i,j) <\epsilon} \,.
\label{eq:NBC_HK}
\end{equation}
 \end{widetext}
The opinion updates between two nodes, $i$ and $j$, can be asymmetric (or even unidirectional) because {the distances} $d_{\text{NBC}}(i,j)$ and $d_{\text{NBC}}(j,i)$ need not be equal even if nodes $i$ and $j$ have equal confidence bounds. In fact, it is typically true that $d_{\text{NBC}}(i,j) \neq d_{\text{NBC}}(j,i)$. Setting $\sigma = 0$ yields a ``pure'' NHK model.


Given a network and an initial opinion profile, the opinion updates in our NHK model are deterministic. Properties of this model include the following:
\begin{itemize}
	\item{The order of the node opinions can change with time.
	That is, $x_i(t) \leq x_j(t)$ for some time $t$ does not imply that $x_i(t + \tau) \leq x_j(t + \tau)$ for all times $\tau > 0$. Switches in node-opinion order can arise from two nodes compromising
	their opinions when their opinions differ by more than the confidence bound (i.e., {when} $|x_i - x_j| > \epsilon$ for a homogeneous confidence bound $\epsilon$) if their neighborhood's mean opinion lies within the confidence bound. 
In the NHK model, it is also possible for a node's opinion to ``jump'' from one opinion cluster to another. Such ``opinion jumping'' also arises in BCMs with polyadic interactions~\cite{hickok2022bounded}, although it occurs for a different reason.}
	\item{If the opinion difference between two nodes does not lie within a confidence bound at a certain time, these nodes can still influence each other through their neighborhoods. Their opinions can subsequently evolve to lie within the confidence bound.}
	\item{The connectedness of a pruned network (i.e., a network that we obtain by removing edges between nodes whose opinions differ by more than the confidence bound)
	is no longer a necessary condition for consensus (as it was in the HK model). It is possible for a pruned network to become disconnected at some time but then become connected again later.}
	\item{Define the ``neighborhood-pruned network" of a network to be the subnetwork that we obtain by removing all edges 
	that are incident to at least one node whose mean neighbor opinion is not within the confidence bound of the other incident node.
		 In the pure NHK model, a necessary condition for consensus is that the neighborhood-pruned network is connected at all times.}
		\item{The opinion-update matrix $W$ is not type-symmetric because neighborhood-based influence is not symmetric (as node $i$ can influence node $j$ even when node $j$ does not influence $i$). That is, $d_{\text{NBC}}(i,j) = 0$ does not imply that $d_{\text{NBC}}(j,i) = 0$. Therefore, the NHK model does not satisfy the conditions in Theorem 2 of \cite{lorenz2005stabilization} for convergence to a nonzero number of opinion clusters with consensus within each opinion cluster.}
\end{itemize}


\subsection{Neighborhood-based network adaptation}
\label{sec: rewiring_opinions}

We now incorporate network adaptation into our NBCMs. In our adaptive NBCMs, nodes incorporate both the opinions of adjacent agents and the mean neighbor opinions of adjacent agents when determining whether to establish or sever an edge. At each discrete time, we choose $f = nN$ edges uniformly at random without replacement. (We take $f = \lceil nN \rceil$ if $nN$ is not an integer.)
For each chosen edge, if the opinions of one or both of its incident nodes are within the confidence bound of each other, then the opinions update according to Eq.~\eqref{eq:NBC_DW}. 
We rewire the edge if
\begin{widetext}
	\begin{align}
		\min\{\sigma d_{\text{BC}}(i,j) + (1 - \sigma) d_{\text{NBC}}(i,j),\sigma d_{\text{BC}}(j,i) + (1 - \sigma) d_{\text{NBC}}(j,i)\} > \zeta \, ,
		\label{eq:zeta}
	\end{align}
\end{widetext}
where $i$ and $j$ are the two nodes that are incident to the edge. (If the inequality \eqref{eq:zeta} holds after updating node opinions, we rewire the edge even if its incident nodes just compromised.) An edge that satisfies the inequality \eqref{eq:zeta} is a ``discordant" edge, so $\zeta \in [0,1]$ is a discordance threshold. A larger value of $\zeta$ entails less tolerance of different opinions and a more stringent requirement to maintain an edge. If $\zeta \geq \epsilon$ {(which is the case for all of our simulations)}, it is only possible to sever an edge if the distance between the opinions of its incident nodes is at least as large as the confidence bound $\epsilon$. When $\zeta = \epsilon$, we rewire any chosen edge in which at least one of the incident nodes does not compromise its opinion. The asymmetry in \eqref{eq:zeta} is important. It is possible to sever the edge between nodes $i$ and $j$ if node $i$'s opinion is sufficiently different from $j$'s opinion, even if node $j$'s opinion is not far enough from $i$'s opinion to exceed the discordance threshold.

When we remove an edge, we select one of its incident nodes, with equal probability of each, to rewire to a new node. Suppose that we select node $i$. Node $i$ considers each node $k$ in the set $K$ of nodes that are not currently in its neighborhood, including the node from which it just detached. We employ one of the following rewiring strategies:
\begin{itemize}
\begin{widetext}
	\item{\textit{Random rewiring}: Node $i$ attaches to a random node (irrespective of the opinion of that node). In this random rewiring strategy, node $i$ chooses a node $k \in K$ uniformly at random. Henceforth, whenever we use the term ``random rewiring", we mean this uniform-at-random rewiring strategy.} 
	\item{\textit{Homophilic rewiring}: Node $i$ attaches to a node $k \in K$ with probability
		\begin{equation}
			P (i \rightarrow k) = \frac{1 - (\sigma  d_{\text{BC}}(i,k) + (1 - \sigma) d_{\text{NBC}}(i,k))}{\sum_k (1 - (\sigma  d_{\text{BC}}(i,k) + (1 - \sigma) d_{\text{NBC}}(i,k)))} \,.
			\label{eq:opn_rewiring}
		\end{equation}	
		We set $P(i \rightarrow i) = 0$ to prevent self-edges. For a progressively larger distance $\tilde{d} := \sigma d_{\text{BC}}(i,k) + (1 - \sigma) d_{\text{NBC}}(i,k))$ between nodes $i$ and $k$, there is a progressively smaller probability that node $i$ rewires to attach to $k$. The neighborhood-tuning parameter $\sigma$ allows us to interpolate between two different types of homophily.
	\begin{itemize}
		\item{Dyadic homophilic rewiring: Node $i$ is more likely to attach to nodes whose opinions are closer to its opinion (as in the rewiring strategy in \cite{kan2023}).}
		\item{Transitive homophilic rewiring: Node $i$ is more likely to attach to nodes with a mean neighbor opinion that is closer to its opinion.}
	\end{itemize}
We consider purely dyadic homophilic rewiring by setting $\sigma = 0$; this captures the idea that people are more likely to befriend somebody that they perceive as similar to themselves. We consider purely transitive homophilic rewiring by setting $\sigma = 1$; this captures the idea that people are more likely to befriend somebody with friends that they perceive as similar to themselves.}
\end{widetext}
\end{itemize}

In Algorithm 1, we outline our algorithm for our NDW model with neighborhood-based network adaptation.

\begin{figure*}
	\begin{framed}
		\caption*{{{\bf Algorithm 1}: Pseudocode for our algorithm to simulate the NDW model with network adaptation.}
		}
		\label{alg-multilayer}
		\begin{algorithmic} 
			\STATE Consider a graph $G = (V,E)$ with a set $V$ of nodes and a set $E$ of edges.
			\STATE \textbf{\# Initialization}
			\STATE Initialize the opinion of each node 
			to a uniformly random value in $[0,1]$.
			\STATE Choose rewire $\in \{\text{random}, \text{homophilic}\}$.
			\FOR {$t <  t_{\text{max}}$ or convergence = False}
			\STATE \hspace{1em} \textbf{\# Opinion spreading}
			\STATE \hspace{1em} Choose $f = nN$ edges.  {\qquad \qquad $\star$} {(If $nN$ is not an integer, then $f = \lceil nN \rceil$.)}
			\STATE \hspace{1em} \textbf{for} each chosen edge $e \in E$: 
			\STATE \hspace{2em} Consider nodes $v_1$ and $v_2$ that are incident to $e$.
			\STATE \hspace{2em} Update the opinions of nodes $v_1$ and $v_2$ using Eq.~\eqref{eq:NBC_DW}.
			\STATE \hspace{2em} \textbf{\# Opinion-driven network adaptation} 
			\STATE \hspace{2em} \textbf{if} $e$ is discordant [see Eq.~\eqref{eq:zeta}]:
			\STATE \hspace{3em} Remove the edge $e$.
			\STATE \hspace{3em} Select a node that is incident to the edge $e$, with equal probability of the two incident nodes. Denote the chosen node by $z$.
			\STATE \hspace{3em} Let $V_z = \{i \in V \, \text{ such that } \, A_{i,z} = 0\}$.
			\STATE \hspace{3em} \textbf{if} rewire = random:
			\STATE \hspace{4em} Choose a node $k \in V_z$ uniformly at random.
			\STATE \hspace{4em} Add an edge so that $A_{k,z} = 1$.
			\STATE \hspace{3em} \textbf{if} rewire = homophilic:
			\STATE \hspace{4em} Choose a node $k \in V_z$ with probability given by Eq.~\eqref{eq:opn_rewiring}. 
			\STATE \hspace{4em} Add an edge so that $A_{k,z} = 1$.
			\ENDFOR
		\end{algorithmic}
	\end{framed}
\end{figure*}

\medskip


\section{Behavior and numerical simulations of the adaptive NDW model}
\label{sec:opinion_empirics}

We conduct numerical simulations to examine the qualitative behavior of our adaptive NDW model.


\subsection{Opinion {states in our adaptive NDW model with homophilic rewiring}}\label{oppy}

In our simulations, we consider the following types of opinion states:
\begin{itemize}
	\item{Consensus: We say that a system is in consensus at a given time if all agents have the same opinion at that time.}
 	\item{Polarization: We say that a system is polarized at a given time if there are two opinion clusters at that time. Our notion of polarization does not consider the sizes of these clusters (no matter how small they are).}
	\item{Fragmentation: We say that a system is fragmented at a given time if there are three or more opinion clusters at that time. We again do not consider the sizes of these clusters.}
	\item{Pseudo-consensus: Consider a set of nodes with at least two distinct opinion values. If the difference in the opinions between any two consecutive nodes (which we order by their opinion values) is less then the confidence bound, then the nodes are in pseudo-consensus. (For example, a set of nodes with opinion values $\{0.1,0.2,0.3\}$ is in pseudo-consensus for $\epsilon > 0.1$.)
		Polarized and fragmented states can have opinion clusters in pseudo-consensus. In our simulations, we choose numerical tolerances to quantify what it means for node opinions to differ from each other. 
		We categorize nodes as having the ``same'' opinion when their opinion values differ by less than $10^{-6}$. In practice, in our computations, all nodes with ``different" opinions differ in their opinion values by at least $10^{-2}$.}
\end{itemize}

Our non-adaptive and adaptive NDW models can have a variety of qualitative dynamics. Examples of such dynamics are the following:
\begin{itemize}
	\item{Convergence: There exists a time $T$ such that the opinion profile $X(T) = X(T + t)$ for all $t \geq 0$.}
	\item{Pseudo-convergence: There exists a time $T$ such that the opinion profile $X(T) = X(T + t)$ for all $t \in [0, \tau)$ but $X(T) \neq X(T + \tau)$.}  
	\item{Temporary consensus: Nodes that have the same opinions at some time but then fall out of consensus at a later time are in a temporary consensus when
	they all have the same opinion.}
	\item{Opinion jumping: An node's opinion changes by a value that is more than the confidence bound $\epsilon$ in a single time step. When a node's opinion jumps, it
	can move from one opinion cluster to another. Opinion jumping can also occur in DW models with polyadic interactions \cite{hickok2022bounded}.}
	\item{Opinion crossing: Consider two nodes $i$ and $j$ with opinions $x_i(t)$ and $x_j(t)$ at time $t$. If $x_i(t) > x_j(t)$ but $x_i(t + \tau) < x_j(t + \tau)$ for some time $\tau > 0$, then the opinions cross each other.	
    Opinion crossing can occur in both our NDW and NHK models. It can also occur in the standard DW model, but it cannot occur in the standard HK model (under the mild assumptions that we stated in Section \ref{sec:BCMs}).}
\end{itemize}


\begin{figure*}[htbp!]
\includegraphics[width=\linewidth]{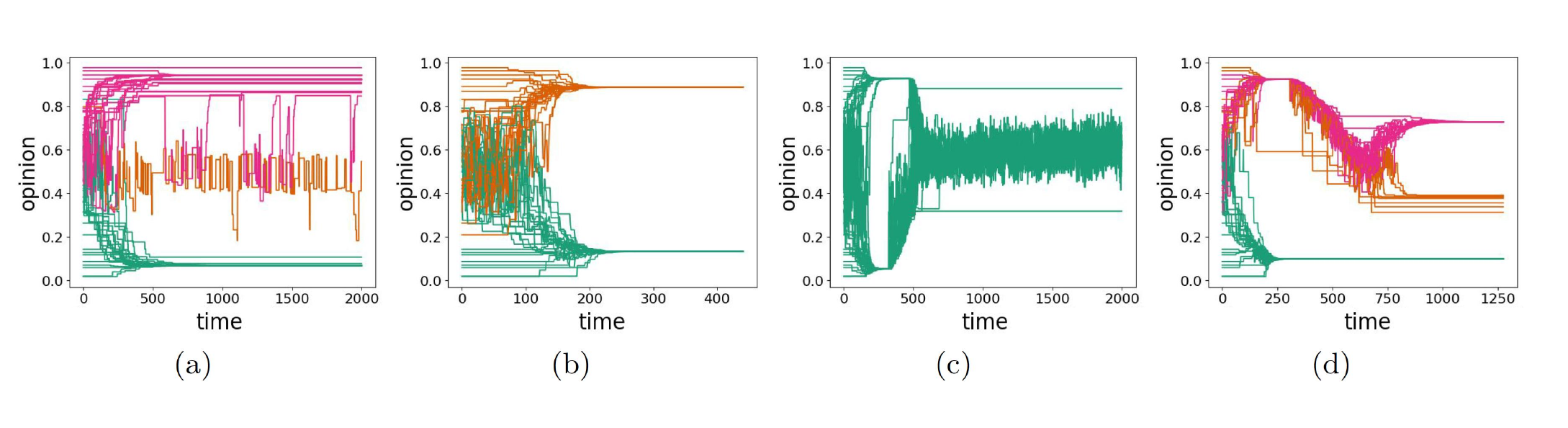}
	\caption{Several examples of opinion dynamics in our neighborhood DW (NDW) model with homophilic rewiring. In each panel, we show one simulation on a network.
		The neighborhood-tuning parameter $\sigma$, the confidence bound $\epsilon$, and the discordance threshold {$\zeta$} have the values
		(a) $\sigma = 0$, $\epsilon = 0.1$, and $\zeta = 0.4$;
		(b) $\sigma = 0.1$, $\epsilon = 0.2$, and $\zeta = 0.2$;
		(c) $\sigma = 0$, $\epsilon = 0.2$, and $\zeta = 0.4$;
		and (d) $\sigma = 0.1$, $\epsilon = 0.1$, and $\zeta = 0.3$.
		At each discrete time, we consider $f = 0.2 N$ dyads, where $N = 100$ is the size (i.e., the number of nodes) of the network. 
		The convergence parameter is $\rho = 0.3$. We choose edges uniformly at random; if an edge is discordant, we rewire it using the homophilic rewiring strategy~\eqref{eq:opn_rewiring}.
		It is possible for the same node to rewire multiple times. In each simulation, the initial network is a $G(N,p)$ Erd\H{o}s--R\'{e}nyi (ER) graph with an independent, homogeneous probability $p = 0.3$ of an edge between each pair of nodes.
		We initialize each node opinion to a uniformly random value in the interval $[0,1]$. All depicted simulations use the same initial network and the same set of initial opinions.
		We terminate a simulation either when it reaches our stopping criterion or when $t_{\text{max}} = 2000$ time steps have elapsed (whichever occurs first).
		We color the opinion trajectories of each node according to its opinion value at the end of a simulation. Any two nodes whose opinions differ by at least the confidence bound $\epsilon$ are in different colors.}
	\label{fig:eg}
\end{figure*}

In Fig.~\ref{fig:eg}, we show examples of various behaviors in the NDW model~\eqref{eq:NBC_DW} {with homophilic rewiring}. We color {the node} opinions according to their final opinion clusters. In each simulation, {the initial network is} a $G(N,p)$ Erd\H{o}s--R\'{e}nyi (ER) graph with $N = 100$ nodes and an independent, homogeneous probability $p = 0.3$ of an edge between each pair of nodes. We initialize the node opinions to uniformly random values in the interval $[0,1]$. {At} each discrete time, we repeat a three-step process for $f = nN$ distinct edges (as before, we take $f = \lceil nN \rceil$ if $nN$ is not an integer),
which we choose uniformly at random without replacement. For each selected edge, we update the opinions of the incident nodes according to the opinion-update rule \eqref{eq:NBC_DW}. If the edge is discordant, we rewire it using one of the strategies in Section~\ref{sec: rewiring_opinions}. (In Fig.~\ref{fig:eg}, we use the homophilic rewiring strategy~\eqref{eq:opn_rewiring}.)
We end a simulation either when it satisfies a stopping criterion or after $t_{\text{max}} = 2000$ time steps (whichever occurs first).
 Our stopping criterion is that no node changes its opinion by more than $10^{-3}$ in any time step for 200 consecutive time steps.

In Fig.~\ref{fig:eg}(a), we show a fragmented state with pseudo-consensus in the pink opinion cluster (top) and green opinion cluster (bottom).
 The orange opinion cluster (middle) has not converged after 2000 time steps. We observe
 opinion jumping at approximately times $600$ and $1900$; some opinion values (in pink) change by more than the confidence bound
 $\epsilon$ in a single time step. In Fig.~\ref{fig:eg}(b), we show an example of polarization. In this simulation, there are also several opinion crossings. 
  In Fig.~\ref{fig:eg}(c), two opinion clusters have converged (to opinions with values of approximately
  $0.9$ and $0.3$) and one opinion cluster has not converged after 2000 time steps. The opinions in this last cluster oscillate wildly between values of about $0.45$ and $0.75$.
  In Fig.~\ref{fig:eg}(d), we observe a variety of interesting behaviors. These behaviors include
  fragmentation into three opinion clusters, pseudo-consensus in the orange opinion cluster (middle), consensus in the pink (top) and green (bottom) opinion clusters, opinion crossing (which is most visible at early times), and opinion jumping (which is most visible in orange). Additionally, the pink opinion cluster has a temporary consensus between the approximate times $250$ and $350$.


\subsection{{Opinion dynamics in our adaptive NDW model with homophilic rewiring}}
\label{sec:op_network}

\begin{figure*}[htbp!]
\includegraphics[width=\linewidth]{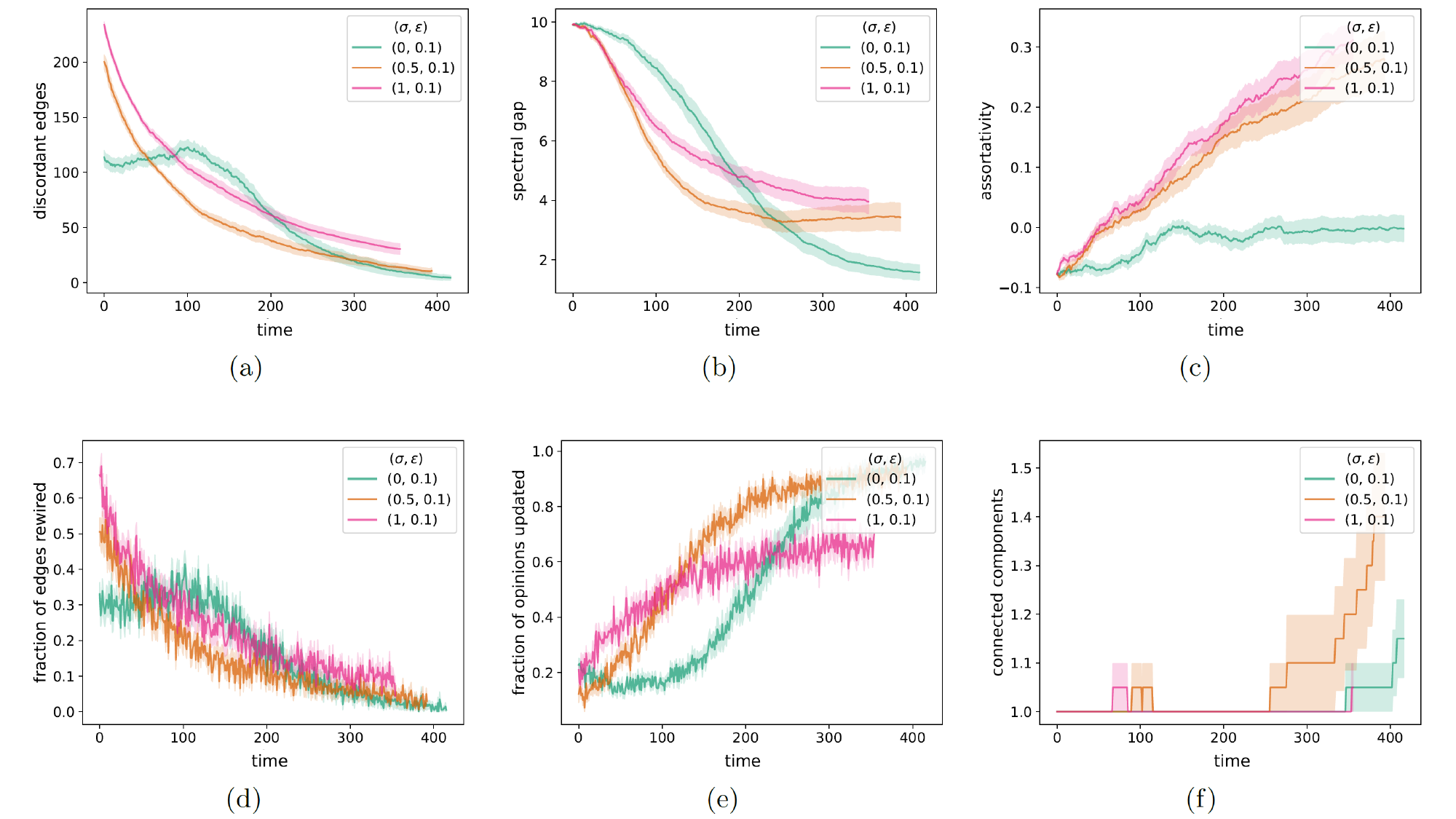}
	\caption{We examine network adaptation as a function of time in our NDW model by plotting (a) the number of discordant edges, (b) the spectral gap of the {associated} adjacency matrix, (c) a degree-assortativity coefficient, (d) the fraction of edges in the chosen dyads that rewire (i.e., that are discordant), (e) the fraction of nodes in the chosen dyads that update their opinions, and (f) the mean number of connected components. We compare the network properties 
		for a pure DW opinion-update rule (i.e., $\sigma = 1$), a mixed DW and NDW opinion-update rule with $\sigma = 0.5$, and a pure NDW opinion-update rule (i.e., $\sigma = 0$).  
		{In each simulation, the initial network is a} $G(N,p)$ ER network with $N = 50$ nodes and an independent, homogeneous probability $p = 0.3$ of an edge between each pair of nodes.
		For each network, we initialize each node opinion to a uniformly random value in $[0,1]$. We plot means of 20 simulations, with the same 20 initial networks and sets of initial opinions for 
		each panel. The shaded regions indicate the standard error. The confidence bound is $\epsilon = 0.1$, the discordance threshold is $\zeta = 0.2$, the number of edges that we choose at
		each discrete time for interaction is $f = 0.2 N$, and the convergence parameter is $\rho = 0.3$.
		We terminate a simulation either when it reaches our stopping criterion or when $t_{\text{max}} = 2000$ time steps have elapsed (whichever occurs first).}
	\label{fig:time-varying}
\end{figure*}

In our adaptive NDW model, the structure of a network can change at each discrete time. These structural changes, in turn, affect the opinion dynamics on the network. 
In Fig.~\ref{fig:time-varying}, we show the time evolution of the network properties in an adaptive network with homophilic rewiring~\eqref{eq:opn_rewiring}.
We compare our {adaptive} NDW model to a baseline {adaptive} DW model for different values of the neighborhood-tuning parameter $\sigma \in [0,1]$, where $\sigma = 1$ corresponds to {pure} DW opinion updates and rewiring that depends only on dyadic homophily and $\sigma = 0$ corresponds to pure NDW opinion updates and rewiring that depends only on transitive homophily. 
As we will illustrate, the network properties depend in an interesting way on $\sigma$. In our simulations, the initial network is a $G(N,p)$ ER graph
with $N = 50$ nodes and an independent, homogeneous probability $p = 0.3$ of an edge between each pair of nodes. We initialize the node opinions to uniformly random values in $[0,1]$.
 We indicate the other parameter values in the caption of Fig.~\ref{fig:time-varying}. Each plot is a mean of 20 simulations with the same parameter values but different randomizations, including both different ER graphs and different sets of initial opinions, with standard errors indicated by the shaded regions. We consider situations that meet our stopping criterion.

In Fig.~\ref{fig:time-varying}(a), we plot the time evolution of the number of discordant edges. We observe a marked difference between the {pure} DW model, in which the number of discordant edges decreases almost monotonically as a function of time, and the pure NDW model, in which the number of discordant edges tends to increase early in simulations (after a small initial dip) and subsequently decreases almost monotonically. In Fig.~\ref{fig:time-varying}(b), we show the time evolution of the spectral gap of the adjacency matrix $A$. The spectral gap is the absolute value of the difference between the two largest eigenvalues of $A$. For a time-independent network, the spectral gap is inversely proportional to the relaxation time of the mixing of a standard random walk on the network (and it is also related to network community structure)~\cite{masuda2017random,mieghem2023}. Intuitively, if nodes consider the neighborhoods of nodes when adapting their opinions, there is an averaging effect that disincentivizes the clustering of opinions into distinct communities. Notably, the behavior of the NDW model with $\sigma = 0.5$, which assigns equal importance to neighborhood influence and dyadic influence, does not simply interpolate between the pure DW model and the pure NDW model.

In Fig.~\ref{fig:time-varying}(c), we plot the time evolution of the degree-assortativity coefficient, which is equal to the Pearson correlation coefficient between the node degrees of adjacent nodes \cite{newman2018}. 
Positive values of degree assortativity indicate a positive correlation between nodes of similar degrees. For all examined values of the neighborhood-tuning parameter $\sigma$ and the confidence bound $\epsilon$, we observe that degree assortativity tends to increase with time, as expected for homophilic rewiring. Additionally, the degree assortativity is much smaller for the pure NDW model (for which $\sigma = 0$) than for the other examined situations. One tends to select nodes with larger degrees for possible opinion updates. Consequently, for simulations that start with uniformly distributed opinions, when we uniformly random select edges for a potential compromise between their incident nodes, we observe that larger-degree nodes' opinions are more likely than smaller-degree nodes' opinions to approach their neighbors' opinions. For the pure DW model (for which $\sigma = 1$), one can thus expect large-degree nodes to often attach to other large-degree nodes, leading to a large degree associativity. This feature is tempered in the NDW model, where neighborhoods also play a role in determining the adjacencies between nodes, so one {can expect the pure NDW model with purely transitive homophilic rewiring to have smaller degree assortativity than the pure DW model with purely dyadic homophilic rewiring.} Once again, the mixed NDW {model} does not simply interpolate between the {pure} DW model and the pure NDW model.

In Fig.~\ref{fig:time-varying}(d), we plot the time evolution of the fraction of the $f$ chosen edges in a time step
that we rewire due to discordance. Recall that we consider $f$ edges (i.e., $f$ dyads) at each discrete time, but we only rewire the discordant edges.
We observe a pattern that resembles the one in Fig.~\ref{fig:time-varying}(a), although Fig.~\ref{fig:time-varying}(d) is much noisier.
In Fig.~\ref{fig:time-varying}(e), we examine the time evolution of the fraction of the nodes in the chosen dyads that update their opinions. It is unsurprising
but still worth remarking that the fraction of edges that are discordant is not inversely correlated with the fraction of opinion updates. Edge discordance and opinion updates are governed by different parameters. Interestingly, although the pure NDW model initially has the fewest opinion updates, it eventually has the most opinion updates. We consider the neighborhood-tuning parameter values $\sigma \in \{0,0.5,1\}$, and we observe noticeable differences in the total {amount that opinions change} for the different values of $\sigma$.

In Fig.~\ref{fig:time-varying}(f), we plot the time evolution of the mean number of connected components of the networks. The pure NDW model and the pure DW model both typically yield the same mean number of connected components (usually $1$), although the pure NDW model eventually produces more components than the pure DW model. Interestingly, we eventually obtain the largest mean number of connected components for $\sigma = 0.5$, which uses a mixture of the NDW and standard DW opinion-update rules.

\subsection{Effect of parameters on opinion clusters in the NDW model with homophilic rewiring}\label{new-c}

We calculate a few different quantities to study the properties of the opinion clusters in the NDW model with homophilic rewiring. We examine how the number of opinion clusters, the relative sizes of the two largest opinion clusters, and the dispersion index~\cite{derrida1986multivalley} change as we vary the confidence bound $\epsilon$, the neighborhood-tuning parameter $\sigma$, and the discordance threshold $\zeta$. We consider situations that meet our stopping criterion. 
 
The dispersion index is~\cite{derrida1986multivalley}
\begin{equation}
	{\Delta} = \frac{\sum_i s_i^2}{\left(\sum_i s_i\right)^2}	\,,
\end{equation}
where $s_i$ is the size of the $i^{\text{th}}$ opinion cluster. An opinion profile with $r$ equal-sized clusters yields $\Delta = 1/r$.

\begin{figure*}[htbp!]
	\includegraphics[width=\linewidth]{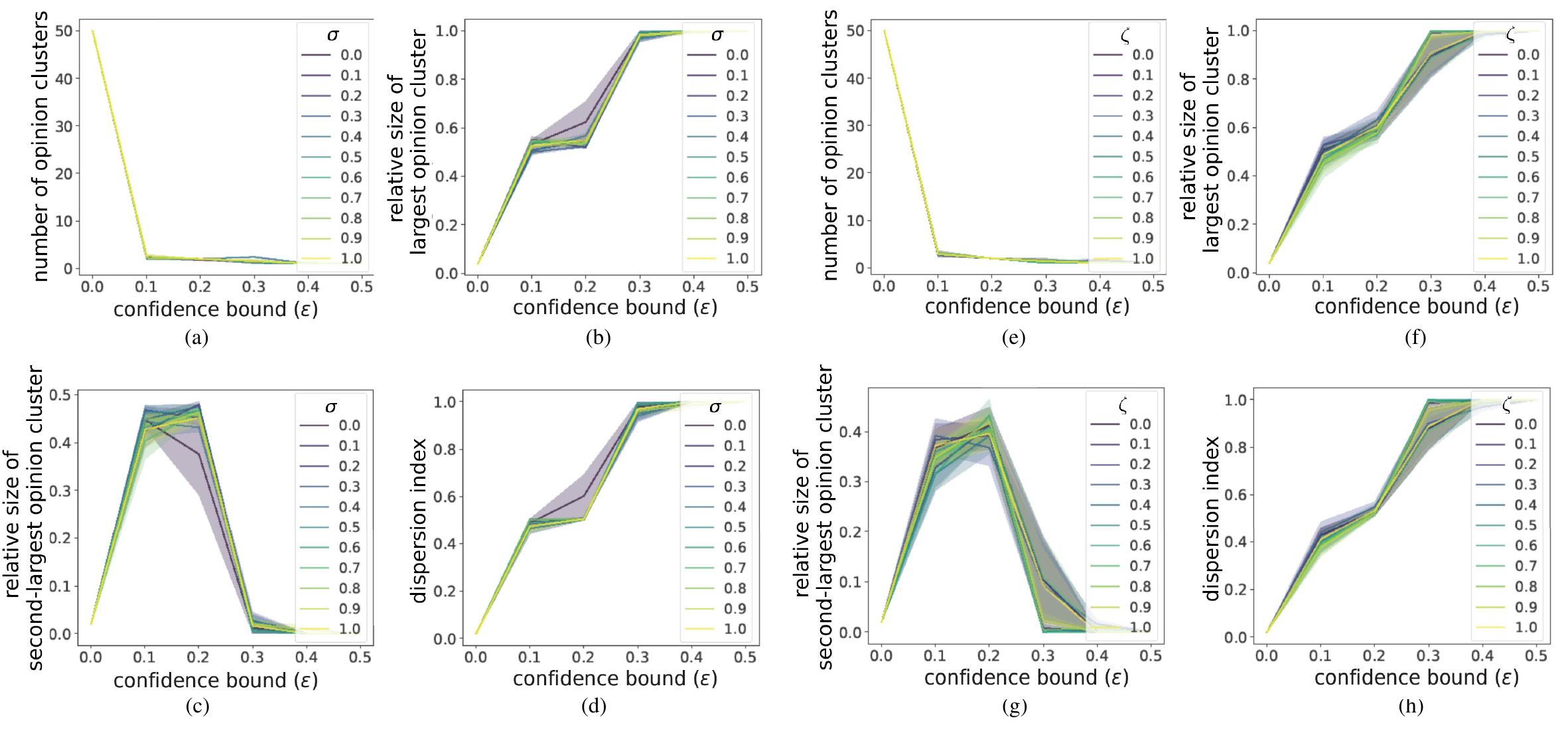}
	\caption{We illustrate the dependence of several final opinion-profile properties in our NDW model with homophilic rewiring on the confidence bound $\epsilon$ for different values of (a, b, c, d) the neighborhood-tuning parameter $\sigma$ and (e, f, g, h) the discordance threshold $\zeta$. We show (a, e) the number of opinion clusters, (b, f) the fraction of nodes in the largest opinion cluster, (c, g) the fraction of nodes in the second-largest opinion cluster, and (d, h) the dispersion index $\Delta$. In each simulation, the initial network is a $G(N,p)$ ER graph with $N = 50$ nodes and an independent, homogeneous probability $p = 0.3$ of an edge between each pair of nodes. In (a, b, c, d), the discordance threshold is $\zeta = 0.2$. In (e, f, g, h), the neighborhood-tuning parameter is $\sigma = 0.5$. For each network, we initialize each node opinion to a uniformly random value in $[0,1]$. The number of edges that we select at each discrete time for agents to interact is $f = 0.2N$, and the convergence parameter is $\rho = 0.3$. We plot means of 5 simulations, with the same 5 initial networks and the same sets of initial opinions for each panel. The shaded regions indicate the standard error.}
	\label{fig:op}
\end{figure*}

In Fig.~\ref{fig:op}, we show the dependence of several final opinion-profile properties on the confidence bound $\epsilon$ for different values of the neighborhood-tuning parameter $\sigma$. In Fig.~\ref{fig:op}(a), we see for all values of $\sigma$ that the number of opinion clusters decreases drastically as we increase the confidence bound from $\epsilon = 0$ to $\epsilon = 0.1$. When $\epsilon = 0$, nodes do not change their opinions. In Figs.~\ref{fig:op}(b, c), we show the relative sizes (with respect to the total number of nodes) of the largest and second-largest opinion clusters. The relative size of the largest opinion cluster tends to grow with $\epsilon$ and appears to reach $1$ for $\epsilon \geq 0.4$ for all values of $\sigma$.
By contrast, the relative size of the second-largest cluster depends nonlinearly on $\epsilon$, with a peak size {between $\epsilon = 0.1$ and $\epsilon = 0.2$}. The dependence on $\sigma$ is less obvious, although the curve corresponding to $\sigma = 0$ (i.e., the pure NDW model) appears to distinctly different from those for the other values of $\sigma$. The sum of the relative sizes of the two largest opinion clusters [see Figs.~\ref{fig:op}(b, c)] is smaller than 1 for some parameter values, so there are more than two opinion clusters in these situations. However, these additional opinion clusters are very small for most parameter values. In Fig.~\ref{fig:op}(d), we observe that the dispersion index $\Delta$ tends to increase with $\epsilon$. We also see that the dispersion index follows a similar trend as the size of the largest opinion cluster.

In Fig.~\ref{fig:op}(e, f, g, h), we observe similar trends with {the confidence bound} $\epsilon$ as we did in {Fig.~\ref{fig:op}(a, b, c, d)}. Additionally, we do not observe a clear dependence on the discordance threshold $\zeta$. A possible reason is that 
$\zeta$ does not directly influence the opinions themselves. Instead, it influences which edges are rewired.

As we can see in Fig.~\ref{fig:op}, the dynamics of our adaptive NDW model depend strongly on the confidence bound $\epsilon$ and depend weakly on the neighborhood-tuning parameter $\sigma$ and the discordance threshold $\zeta$. The dependence on $\epsilon$ is familiar from studies of ordinary BCMs \cite{bernardo2024,meng2018}. For small values of $\epsilon$, the opinions of neighboring nodes are unlikely to lie within the confidence bound, so nodes are unlikely to influence each other. This tends to yield many opinion clusters. As we increase $\epsilon$, more nodes influence each other and we obtain fewer opinion clusters. The observed nonmonotonic dependence on $\epsilon$ of the sizes of the largest and second-largest opinion clusters [see Fig.~\ref{fig:op}(b, d, f, h)] is interesting and worthy of further investigation.


\subsection{Simulations of our adaptive NDW model with homophilic rewiring on different types of initial networks}\label{new-d}

We now simulate our adaptive NDW model on several additional types of initial networks. These simulations complement our previous simulations on ER graphs. We consider situations that {meet} our stopping criterion and examine the following types of networks:
\begin{itemize}
	\item{Holme--Kim power-law cluster graphs: A Holme--Kim graph \cite{holme2002vertex} is a generalization of the standard Barab\'asi--Albert preferential-attachment graph \cite{newman2018,barabasi1999emergence} that also adds triangles. 
One starts with a set of isolated nodes. In each preferential-attachment step, one adds a new node and connects it to $m$ existing nodes with probabilities that are proportional to their degrees. With probability $\tilde{p}$, for each new node and new edge from the preferential-attachment step, one adds another edge and forms a triangle by connecting the new node to a neighbor of the previously linked node. (If no such new edges are possible, one instead performs another preferential-attachment step.) By adding triangles, one increases the triadic clustering in the network.}
	\item{Newman--Watts--Strogatz (NWS) small-world graphs~\cite{newman1999renormalization}: To create an NWS graph, we start with a ring network. We then connect each node of the ring to its $r$ nearest neighbors, where $r$ is an even positive integer. We then add ``shortcut'' edges as follows. For each edge of the graph, with probability $\tilde{p}$, we add a new edge from one of its incident nodes {(which we choose uniformly at random)} to a uniformly randomly chosen node.} 
	\item{The Zachary Karate Club (ZKC) graph~\cite{zachary1977}: The ZKC graph is an astoundingly popular 34-node social network of friendships in a karate club. This real-world network has well-studied community structure~\cite{girvan2002}.}
\end{itemize}

\begin{figure*}[htbp!]
\includegraphics[width=\linewidth]{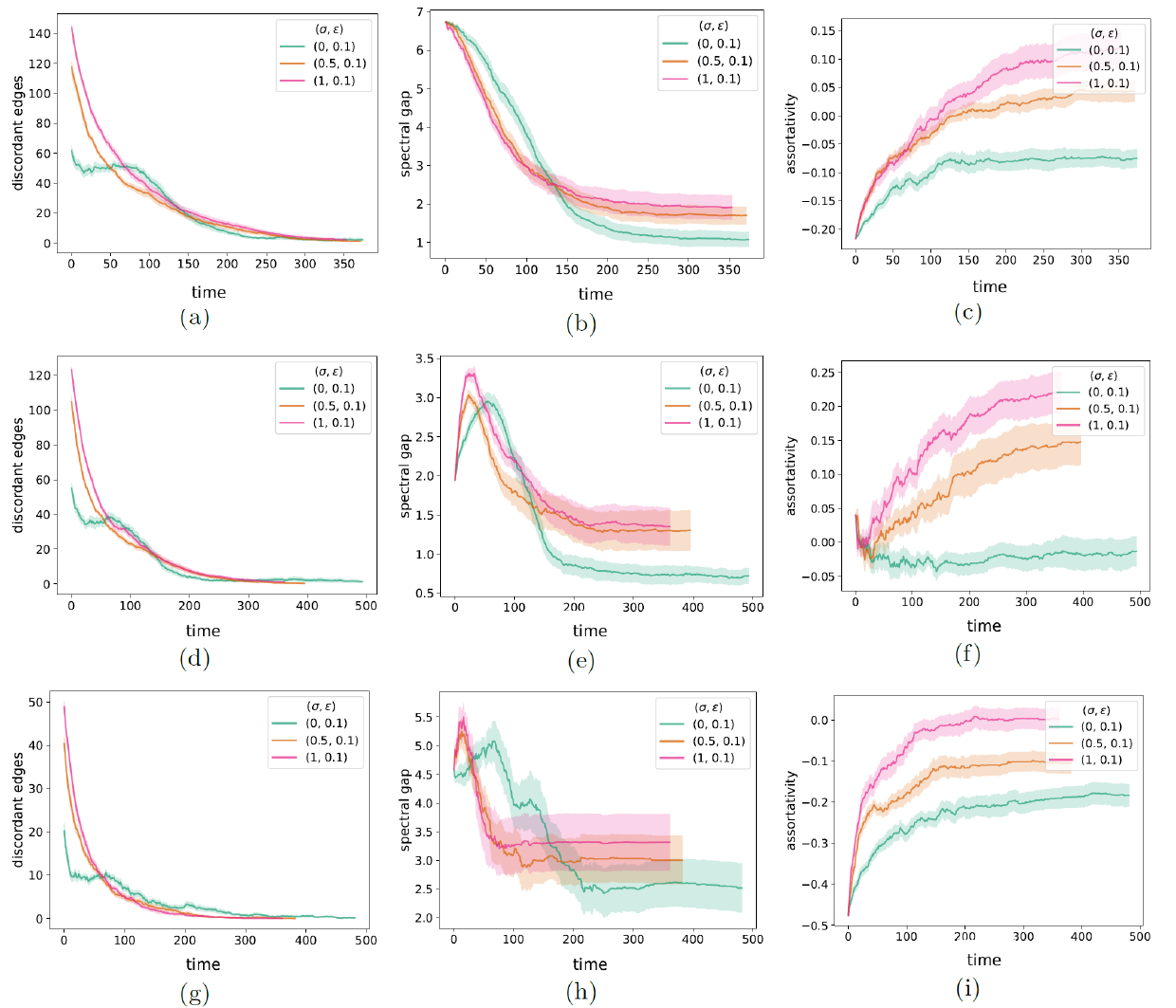}
	\caption{We {simulate} our adaptive NDW model with homophilic rewiring on three types of networks. We consider (top) a Holme--Kim
	power-law graph with clustering, (middle) a Newman--Watts--Strogatz (NWS) small-world graph, and (bottom) the Zachary Karate Club graph.
	For each type of graph, we plot (a, d, g) the number of discordant edges in the network, (b, e, h) the spectral gap of the network's adjacency matrix, and (c, f, i) a degree-assortativity coefficient.
	We compare the network properties for the baseline adaptive DW model (i.e., the neighborhood-tuning parameter is $\sigma = 1$), a mixed adaptive NDW model (with $\sigma = 0.5$), 
	and a pure adaptive NDW (i.e., $\sigma = 0$). For the Holme--Kim graph, we start with 5 isolated nodes, add a new node to attach to 5 existing nodes using linear preferential attachment, and incorporate additional edges for triadic closure with probability $\tilde{p} = 0.3$. For the NWS small-world graph, each node is adjacent to $6$ nearest neighbors and the probability of adding a new edge for each edge is $\tilde{p} = 0.3$.
	The Holme--Kim and NWS graphs have $N = 50$ nodes, and the Zachary Karate {Club graph has $N = 34$ nodes. We initialize each node opinion to a uniformly random value in the interval $[0,1]$.
	Each curve in each panel is the mean of 5 simulations with different initial opinions. We use the same Holme--Kim graph and the same NWS graph for all simulations.}
	 The shaded regions indicate the standard error. The confidence bound is $\epsilon = 0.1$, the discordance threshold is $\zeta = 0.2$, the number of edges that we choose at each discrete time for interaction is $f = 0.2 N$, and the convergence parameter is $\rho = 0.3$. We terminate a simulation either when it reaches our stopping criterion or when $t_{\text{max}} = 2000$ time steps have elapsed (whichever occurs first).}
	\label{fig:different_networks}
\end{figure*}

In Fig.~\ref{fig:different_networks}, we show the results of our simulations of our adaptive NDW model with homophilic rewiring on the different networks. {As one can see in Fig.~\ref{fig:different_networks}(a, d, g) and Fig.~\ref{fig:time-varying}(a), the time evolution of the number of discordant edges is qualitatively similar for all of the examined networks. The} number of discordant edges decreases with time for all examined networks, and the pure adaptive NDW model
(i.e., with $\sigma = 0$) appears to have a small plateau-like structure at early times. Additionally, the spectral gap decreases most rapidly for our pure adaptive NDW model [see Fig.~\ref{fig:different_networks}(b, e, h) and Fig.~\ref{fig:time-varying}(b)].
The spectral gap for the Holme--Kim graph decreases with time, whereas the spectral gaps for the NWS and ZKC {graphs} increase initially before they too tend to decay. The spectral gap for the ZKC graph has additional structure, perhaps due to its strong community structure. Finally, as one can see in Fig.~\ref{fig:different_networks}(c, f, i) and Fig.~\ref{fig:time-varying}(c), the degree assortativity of our adaptive NDW model plateaus at the smallest value, followed by the baseline DW model, and then the mixed NDW model. The approach to the plateau appears to be qualitatively different for the NWS graph than for the other graphs. 


\subsection{Effect of network size for ER graphs} \label{new-e}

We now study the effect of network size on the qualitative dynamics of our adaptive NDW model. We simulate {the adaptive NDW model with homophilic rewiring} on $G(N,p)$ ER networks with {connection probability} $p = 0.3$ and $N = 100$, $N = 200$, and $N = 400$ nodes. Unsurprisingly, we observe that larger networks take longer to reach a steady state. In our simulations with $N = 50$ nodes (see Fig.~\ref{fig:time-varying}), we observe convergence (up to a $10^-3$ error tolerance), with simulations achieving our stopping criterion before our simulations run for $t_{\text{max}} = 2000$ time steps. However, for $N = 400$ nodes, our simulations of the adaptive NDW model with purely transitive homophilic rewiring (i.e., with $\sigma = 0$) run for $t_{\text{max}} = 2000$ time steps without achieving our stopping criterion.

\begin{figure*}[htbp!]
\includegraphics[width=\linewidth]{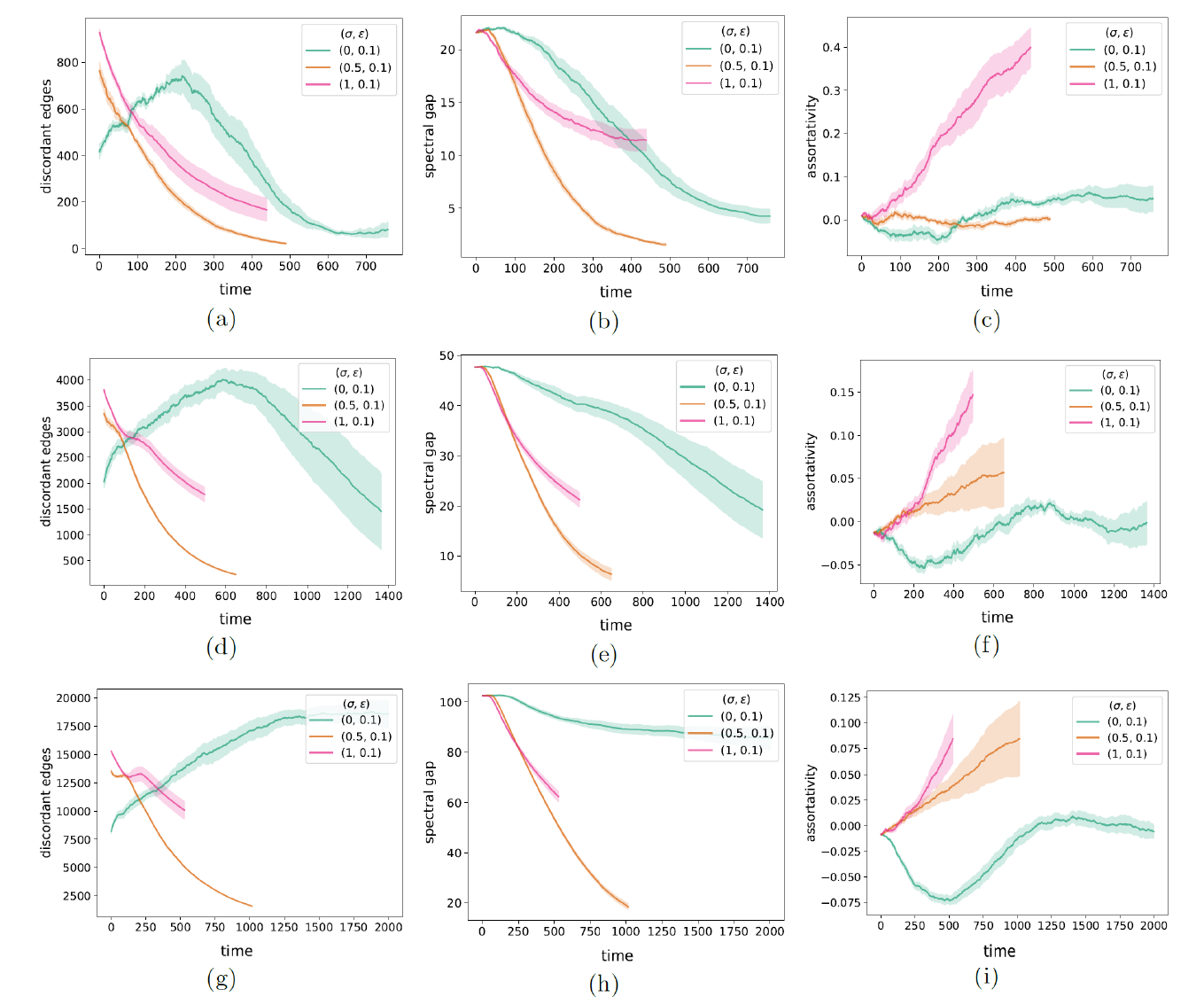}
	\caption{We simulate our adaptive NDW model with homophilic rewiring on $G(N,p)$ ER
	graphs with: (top) $N = 100$ nodes, (middle) $N = 200$ nodes, and (bottom) $N = 400$ nodes. 
	For each of these network families, we plot (a, d, g) the number of discordant edges, (b, e, h) the spectral gap of the adjacency matrix, and (c, f, i) a degree-assortativity coefficient.
	We compare the network properties for a pure DW model with purely dyadic homophilic rewiring (i.e., $\sigma = 1$), a mixed {NDW model (specifically, with $\sigma = 0.5$), and a pure NDW model with purely transitive homophilic rewiring (i.e., with $\sigma = 0$).} In each simulation, the initial network is a $G(N,p)$ ER network (with $N = 100$, $N = 200$, and $N = 400$ nodes in the top, middle, and bottom rows, respectively) and an independent, homogeneous probability $p = 0.3$ of an edge between each pair of nodes. We initialize each node opinion to a uniformly random value in $[0,1]$. We plot means of 5 simulations, with the same 5 initial networks and sets of initial opinions for each panel. The shaded regions indicate the standard error. The confidence bound is $\epsilon = 0.1$, the discordance threshold is $\zeta = 0.2$, the number of edges that we choose at each discrete time for interaction is $f = 0.2 N$, and the convergence parameter is $\rho = 0.3$. We terminate a simulation either when it reaches our stopping criterion or when $t_{\text{max}} = 2000$ time steps have elapsed (whichever occurs first).}
	\label{fig:varying_N}
\end{figure*}

We briefly summarize a few qualitative features of Fig.~\ref{fig:varying_N}. In the left column of Fig.~\ref{fig:varying_N}, we observe that the number of discordant edges in our adaptive NDW model with purely transitive homophilic rewiring (i.e., with $\sigma = 0$) increases with time at first and then eventually decreases. By contrast, for $\sigma = 1$ (i.e., for the adaptive DW model with purely dyadic homophilic rewiring), the number of discordant edges consistently decreases with time. In the middle column of Fig.~\ref{fig:varying_N}, we observe that the spectral gap decreases much more sharply for the the pure DW model (with $\sigma = 1$) and the mixed NDW model ($\sigma = 0.5$) than for the pure NDW model (with $\sigma = 0$). Finally, in the right column of Fig.~\ref{fig:varying_N}, we observe that the pure NDW model with purely transitive homophilic rewiring consistently has the smallest degree assortativity of the three examined situations.

\begin{figure*}[htbp!]
		\includegraphics[width=0.8\linewidth]{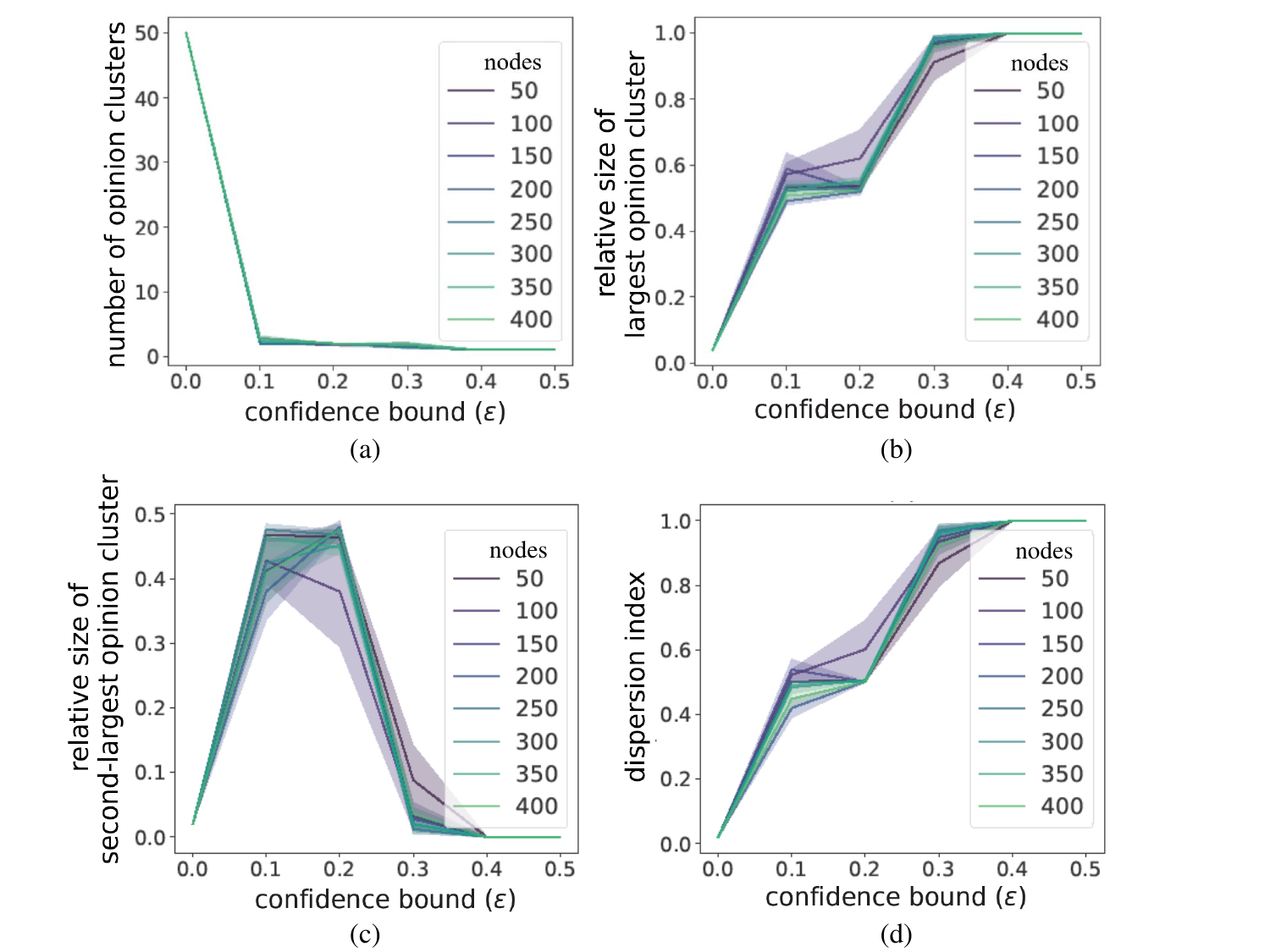}
	\caption{We illustrate the dependence of several final opinion-profile properties in our adaptive NDW model with homophilic rewiring on the confidence bound $\epsilon$ for different values of the network size $N$. 
	We show (a) the number of opinion clusters, (b) the fraction of nodes in the largest opinion cluster, (c) the fraction of nodes in the second-largest opinion cluster, and (d) the dispersion index $\Delta$. 
	{In each simulation, the initial network is a} $G(N,p)$ ER graph with $N$ nodes and an independent, homogeneous probability $p = 0.3$ of an edge between each pair of nodes. For each network, we initialize each node opinion to a uniformly random value in $[0,1]$. The neighborhood-tuning parameter is $\sigma = 0.5$, the discordance threshold is $\zeta = 0.2$, the number of edges that we select at each discrete time for agents to interact is $f = 0.2N$, and the convergence parameter is $\rho = 0.3$. We plot means of 5 simulations, with the same 5 initial networks and sets of initial opinions for each panel. The shaded regions indicate the standard error.}
	\label{fig:changing_nodes}
\end{figure*}

In Fig.~\ref{fig:changing_nodes}, we show the dependence of opinion clusters on the confidence bound $\epsilon$ for different network sizes. We observe that the relative size of opinion clusters (as a fraction of the total network size) does not depend strongly on the number of nodes in the network.

\section{Conclusions and discussion}\label{conclusions}

We studied adaptive bounded-confidence models (BCMs) of opinion dynamics that incorporate neighborhood effects into both opinion dynamics and network adaptation. In addition to the usual dyadic influence of BCMs, these neighborhood BCMs (NBCMs) incorporate transitive influence in determining whether or not agents compromise their opinions when they interact with other agents. In such transitive influence, agents seek the mean opinions of the neighbors of their neighbors (e.g., the friends of friends) to be sufficiently similar to their opinions. They thereby incorporate a notion of ``you are who you know" into opinion updates. We formulated neighborhood-informed generalizations of both the Hegselmann--Krause (HK) model and the Deffuant--Weisbuch (DW) {model, and we argued that the neighborhood HK (NHK) model includes qualitative behavior (such as changes in the order of opinions with time) that cannot occur in the standard HK model.} 

We then developed adaptive NDW and NHK models with neighborhood-informed network adaptation, and we examined the qualitative behavior of the adaptive NDW model in various scenarios. Our neighborhood-informed homophilic rewiring strategy is based on transitive homophily and yields interesting network properties. In our adaptive NDW model, we obtained a smaller degree assortativity, a smaller spectral gap, and fewer connected components than in our baseline {adaptive} DW model. In the adaptive NDW model, we also observed nonmonotonic behavior in the number of discordant edges as a function of time.
	
Neighborhood-based transitivity --- which can arise through both transitive homophily and transitive influence --- exerts notable effects on human behavior~\cite{liu2011trust,christakis2009connected,christakis2013,backstrom2011}. 
It is thus important to incorporate such ideas into models of opinion dynamics. However, as with all other models of opinion dynamics, our NBCMs have several limitations. In particular, we made many simplistic assumptions about human behavior. For example, all of the agents in our models are identical and their confidence bounds are homogeneous. It is relevant to explore heterogeneities in these and other features. 

There are a variety of ways to extend our NBCMs. These extensions include both commonly noted possibilities (such as the incorporation of multidimensional opinions, heterogeneous confidence bounds, and polyadic interactions) and generalizations that align specifically with our model's neighborhood focus. In particular, we considered only the two-step neighbors of agents (e.g., friends of friends), and it will be interesting to examine generalizations of our NBCMs that include more expansive neighborhoods. Additionally, it is worthwhile to extend other types of opinion models to explore neighborhood effects, and our NBCMs provide an illustrative example to help guide such efforts. Another interesting direction is to combine our NBCMs with models of disease spread to explore how {neighborhood-based} opinion dynamics and disease dynamics affect each other.


\section*{Code availability}

Our code is available at \url{https://bitbucket.org/neighborhood-bounded-confidence-model-of-opinion-dynamics}.


\section*{Acknowledgements}

MAP acknowledges financial support from the National Science Foundation (grant number 1922952) through the Algorithms for Threat Detection (ATD) program. SK acknowledges support from the UC Presidential Postdoctoral Fellowship. We thank two anonymous referees for their many helpful comments.





\end{document}